\documentclass{kluwer}    % Specifies the document style.

\newdisplay{guess}{Conjecture}

\begin{document}                                                                                   
\begin{article}
\begin{opening}         
\title{From old Globular Clusters to early Structures in the Universe: the
Formation of old, metal-poor Halos around early-type Galaxies
%\thanks{Footnote to the title with the `thanks' command.}
}
\author{Markus \surname{Kissler-Patig}}  
\runningauthor{Markus Kissler-Patig}
\runningtitle{Old, metal-poor halos around early-type galaxies}
\institute{European Southern Observatory, Garching, Germany}
%, Karl-Schwarzschild-Str.~2, 85748 Garching, Germany}
\date{December 1, 2001}

\begin{abstract}
Old, metal-poor globular clusters trace the formation and evolution of
early-type galaxies. Their are the best probes, at low redshift, of the 
building-up of galaxy halos at high redshift. Their properties constrain
the characteristics of their progenitors. Recent results suggest that DLAs at
$z>3$ are the likely hosts for their formation. Finally, they shed light
on the old, metal-poor halos probably present around all early-type galaxies.
\end{abstract}

\end{opening}           

\section{Why old, metal-poor globular clusters?} 

Old, metal-poor globular clusters are excellent tracers for the
formation of galaxies (see recent reviews by Ashman \& Zepf 1998, Kissler-Patig
2000, Harris 2001). They are present in large numbers around 
{\em all} giant galaxies studied to date and can therefore be used as a common
tool. They are old ($\sim 12\pm2$
Gyr) and thus witnessed the past since a time corresponding to a high redshift 
($z>3$, somewhat dependent on the cosmology). They form as stars
form, i.e.~understanding their epochs of formation allows to study the
star formation history of their host galaxy. Finally, they are simple
stellar populations, easier to model and to understand than the diffuse
stellar light of their host galaxies.

Globular cluster {\em sub-populations} turned out to be very common (Zepf \&
Ashman 1993, and e.g.~Gebhardt \& Kissler-Patig 1999). To first
approximation, a typical globular cluster system shows two broad
sub-populations. An old-metal poor sub-population is almost always present, 
as well as a 
more metal-rich (old to intermediate age, possibly itself divided into
sub-populations). The properties of the latter resemble those of the
diffuse stellar light. Indeed, the integrated light of early-type
galaxies appear to be dominated by the metal-rich component. {\em Thus, the
old, metal-poor globular clusters are a unique opportunity to study the halo 
component of early-type galaxies otherwise difficult to observe}. At the same 
time, they allow to study (at low-redshift) the building up (at high redshift) 
of early-type galaxies.

\section{The properties of the old, metal-poor globular cluster sub-populations}

Only since the mid-90s, studies investigate the properties of the 
various {\em sub-populations} of globular clusters around early-type galaxies, 
as opposed to focusing on the general properties of the entire system.

The first difference noticed (in NGC 4472) between the red and the blue 
sub-populations was their different {\bf density distributions} (Geisler, Lee \& 
Kim 1996). The red clusters are significantly more concentrated towards
the centre than the blue ones. This is found to be true in all other
galaxies studied to date. Furthermore, {\bf the spatial distributions} also
appear to differ. E.g.~in NGC 1380, an S0 galaxy, Kissler-Patig et al.~(1997) 
found the red clusters to follow the ellipticity of the diffuse light while 
the blue clusters where spherically distributed. Finally, {\bf the kinematics}
of the two sub-populations were found to differ systematically in the
studied cases (e.g.~Kissler-Patig \& Gebhardt 1998, Zepf et al.~2000,
C\^ot\'e et al.~2001). In summary, the blue and the red globular
clusters form two distinct sub-populations from their properties, {\em
and the blue (metal-poor) globular clusters have `halo' properties.}

The metal-poor globular cluster also stand out with respect to {\bf their
sizes}. While the sizes of clusters in nearby galaxies were all found to
roughly correspond to the observed sizes in the Milky Way, the metal-poor
clusters appear systematically larger than the metal-rich ones 
at all galacto-centric radii
(e.g.~Kundu \& Whitmore 1998, Puzia et al.~1999, Larsen et al.~2001).
This result is interpreted as a imprint from the time of formation,
hinting at the formation of metal-poor clusters in shallower potential
wells, i.e.~smaller fragments.

{\bf The abundances and abundance ratios} of old, metal-poor clusters
are best studied using spectroscopy of clusters in nearby galaxies. All
old-metal poor clusters are found to lie in the same region not only
in H$\beta$ -- MgFe diagrams, but also in Fe -- Mg diagrams
(e.g.~Kissler-Patig et al.~1998, Cohen et al.~1998, Schroder et al.~2001). 
This points to very similar ages and $\alpha$-element ratios among all old,
metal-poor clusters in the nearby universe, including the Milky Way, M31
and M81.

Consequently, {\bf the mean metalicity} of the metal-poor globular
cluster sub-population in
a given galaxy was found to be constant over a large range of galaxy
sizes, morphologies and metalicity (e.g.~Ashman \& Bird 1993,
Burgarella et al.~2001). The properties of the metal-poor sub-populations 
correlate only weakly (if at all) with their host-galaxy properties.

Finally, the globular cluster luminosity function was show to be a useful
distance indicator, especially when using only old-metal clusters
(e.g.~Kissler-Patig 2000). This, in turn, proves that {\bf the mass 
distribution} of
the globular clusters is universal and independent of the galaxy properties: 
again suggesting a formation process largely unrelated to the final host galaxy.

The conclusion from the above is that {\em old, metal-poor globular
clusters are present in all observed early-type galaxies and have
universal properties that do not depend (or only weakly) on the host
galaxy properties}. This suggests their formation in small fragments,
largely independent of the final host galaxy.

\section{On the nature of their progenitor fragments}

The properties of old-metal poor globular clusters can constrain the
nature of their formation sites. This was discussed by Burgarella, 
Kissler-Patig \& Buat (2001), who identified damped lyman-$\alpha$
systems (DLAs) as likely sites for
the formation of the old, metal-poor globular clusters.

Updating their result, the ages of old, metal-poor globular clusters
(around $12\pm2$ Gyr) correspond to a redshift of formation of
$z=4^{+\infty}_{-1.5}$ in
the current standard cosmology ($\Omega = 0.3, \Lambda
=0.7,$H$_0=70$ km$\cdot$s$^{-1}\cdot$Mpc$^{-1}$). Comparing this with the
latest abundance measurements of DLAs (e.g.~Dessauges-Zavadsky et
al.~2001), indicates that the mean abundance of DLAs at $z>3$ matches
well the $\sim 1/50$ solar mean abundance of the old, metal-poor
globular clusters. These DLAs are thus likely to be the fragments in
which (one or more) globular clusters formed, while DLAs at lower redshift might, 
at least partly, belong to a different category of objects (more evolved 
spirals?). 

From the properties of the old, metal-poor globular clusters we 
learn the following on the nature of the fragments:

$\bullet$ The fragments had metalicities between $-1.0<$[Fe/H$<-2.5$ dex, 
the range spanned by the old, metal-poor clusters.

$\bullet$ Their mass distribution is likely to have been a power-law of slope
around $-2$, transmitting this characteristic imprint to the mass function
of the globular clusters (up to masses around $10^8$ M$_\odot$). The
high mass cut-off as judged from the largest dwarf galaxies hosting
metal-poor globular clusters only, must have lied around
$10^9$--$10^{10}$M$_\odot$.

$\bullet$ The fragments collapse at roughly $z>3$ without suffering the
influence of the final host galaxies of which they formed the halos
later. This way, they probably enriched significantly the intra-galactic
medium.

\section{Where are the old, metal-poor stars in ellipticals?}

Clearly, the presence of old, metal-poor globular cluster sub-populations calls for
a stellar counter-part. Where are the stars associated with this halo
population in early-type galaxies?

There is good evidence that the old, metal-poor globular clusters formed
with a very high specific frequency (S$_N\sim20$), i.e.~have only few
stars associated with them, while the metal-rich globular clusters
formed together with the vast majority of the stars (low S$_N\sim1$--2).

This is supported {\it i)} by the high specific frequency observed in dwarf
galaxies and halos, dominated by old, metal-poor globular clusters, {\it
ii)} by the relatively small amount of metal-poor stars contributing to
the integrated light of early-type galaxies (see Maraston \& Thomas
2000, Lotz et al.~2000), {\it iii)} by the direct comparison of
the metal-poor to metal-rich number ratios for stars and globular
clusters in NGC 5128, the nearest giant elliptical (Harris, Harris \&
Poole 1999): While the metal-rich stars vastly dominate in number the
stellar population, metal-poor and metal-rich globular clusters appear
in roughly similar numbers.

Therefore, we expect only a small population of old, metal-poor stars
associated with the old, metal-poor globular clusters. Such a stellar
population is difficult to detect within the diffuse stellar light
dominated by the metal-rich stars.

\section{The formation of halos around early-type galaxies}

In summary: old, metal-poor globular clusters form an independent `halo' 
sub-population present in all galaxies. Their properties suggest a
formation in small fragments identified with DLAs at $z>3$. {\em This
implies that the metal-poor halos of giant ellipticals formed through
the assembly of (mostly) collapsed small fragments}. Their properties
correlate only weakly if at all with the host galaxy properties,
suggesting that {\em some fragments might have been influence by the
final host galaxy, but many (most?) formed as independent satellites.}

\acknowledgements
Many thanks go to my collaborators D.Burgarella, V.Buat, C.Maraston and
D.Thomas for fruitful discussions on old, metal-poor clusters, stellar
populations and halos.

\end{article}
\end{document}